\documentclass[prl,twocolumn,showpacs,amsmath,amssymb]{revtex4}
\usepackage{pifont,amssymb,epsf,graphicx}
\begin{document}

\title{\bf Local Density of States and Angle-Resolved Photoemission Spectral Function
of an Inhomogeneous D-wave Superconductor}

\author{Ming Cheng }
\author{W. P. Su}

\address{Department of Physics and Texas Center for Superconductivity,
University of Houston, Houston, Texas 77204, USA
}

\date{\today}
\begin{abstract}

  Nanoscale inhomogeneity seems to be a central feature of the
d-wave superconductivity in the cuprates. Such a feature can strongly
affect the local density of states (LDOS)
and the spectral weight functions. Within
the Bogoliubov-de Gennes formalism we examine various inhomogeneous
configurations of the superconducting order parameter to  see which
ones better agree with
the experimental data. Nanoscale large amplitude
oscillations in the order parameter seem to fit the  LDOS data for the
underdoped cuprates. The  one-particle 
spectral function for a general inhomogeneous
configuration exhibits a coherent peak in the nodal direction.
In contrast, the spectral function in the antinodal region is easily
rendered incoherent by the inhomogeneity. This throws  new light
on the dichotomy between the nodal and antinodal quasiparticles in the
underdoped cuprates.
\end{abstract}

\pacs{74.20.-z, 74.72.-h, 74.25.Bt}

\maketitle
\section{I. Introduction}
The scanning tunneling microscopy (STM) and
 the angle-resolved photoemission spectroscopy (ARPES)
are two of the most important tools for unraveling the mystery
of the high-temperature superconductors (HTS).
The existence of a Fermi surface and the d-wave symmetry of the
superconducting state are very important properties revealed by
ARPES~\cite{shen}
in momentum space. In contrast, STM has provided important
complementary information in real space through the measurement
of the local density of states (LDOS). A surprising feature of
the HTS seen through STM is the conspicuous 
inhomogeneity~\cite{pan,kap,lan}.

While some of the STM and ARPES data are straightforward to interpret,
others are not. The inhomogeneity underlies much of the difficulty as
there has not been much theoretical work addressing the effect of
inhomogeneous d-wave superconductivity (DSC) 
on the LDOS~\cite{wan,anna,atk} and the one-particle
spectral function~\cite{bish,deg,hal}.
 This paper is intended to partially remedy the
situation.

In the absence of a complete theory of HTS, what we have done is to
examine various types of inhomogeneity for comparison with the STM
and the ARPES data. In this way we hope to extract as much information
from the data as possible. In the following, we first introduce the
model Hamiltonian and the method of calculation. The calculated 
results for LDOS and the spectral function are presented and their
implications are discussed.

\section{II. Model Hamiltonian}

In this work, we focus exclusively on the
effect of inhomogeneous d-wave pairing field. We therefore adopt
the following Hamiltonian

\begin{eqnarray}
H=\sum_{k\sigma}(\epsilon_k-\mu)c_{k\sigma}^{\dagger}c_{k\sigma}+
\sum_{<i,j>}(\Delta_{i,j}c_{i\uparrow}c_{j\downarrow}+H.c.)
\end{eqnarray}

where $c_{i\sigma}^{\dagger}$ creates an electron on site i with spin 
$\sigma$, and $c_{k\sigma}^{\dagger}$ an electron with momentum k and
spin $\sigma$.
$<i,j>$
is a nearest neighbor pair. 
The kinetic energy is give by

\vspace{0.1in}

\begin{widetext}
$$\epsilon_k=t_1(cosk_x+cosk_y)/2+t_2cosk_xcosk_y+
t_3(cos2k_x+cos2k_y)/2+t_4(cos2k_xcosk_y+cosk_xcos2k_y)/2+t_5cos2k_xcos2k_y,
$$
\end{widetext}

\vspace{0.1in}

where the hopping parameters 
 $t_{1-5}$=-0.5951, 0.1636, 
-0.0519, -0.1117, 0.0510 eV are from $Bi_2Sr_2CaCu_2O_{8+\delta}$ band structure determined by Norman et al.~\cite{nor}.
 $\Delta_{i,j}$ is the d-wave pairing amplitude over a
nearest neighbor pair $<i,j>$.

For any spatial distribution of $\Delta_{ij}$, the Hamiltonian (1) can
in principle be straightforwardly diagonalized after performing a
Bogoliubov transformation. In reality, due to the large lattice size
involved we need to resort to a special technique for calculating the
LDOS.

 Gagliano and Balseiro~\cite{gag} have proposed an efficient method for
calculating the resolvent $G_{A}=<\psi_{0}|A^{\dag}(Z-H)^{-1}A|\psi_{0}>$.
$G_{A}(Z)$ is expressible as a continued fraction~\cite{hay,gros}

 $$G_{A}(Z)=\frac{<\psi_{0}|A^{\dag}A|\psi_{0}>}{Z-a_{0}
-\frac{b_{1}^{2}}{Z-a_{1}-\frac{b_{2}^{2}}{Z-\cdot\cdot\cdot}
}}.
\eqno{(2)}$$

where the coefficients $a_{i}$ and $b_{i}$ can be obtained from $A|\psi_{0}>$
by repeated application of the Hamiltonian $H$. The Fourier transform
of the self-correlation function $C_{A}(t-t^{'})=<\psi_{0}|A^{\dag}(t)
A(t^{'})|\psi_{0}>$
 can be recovered from the imaginary part
of the resolvent $C_{A}(\omega)=\frac{1}{\pi}ImG_{A}(\omega+i\eta+E_{0})$.

The above method is originally devised for calculating the dynamical 
properties of quantum many-body systems~\cite{gag}, 
but it can be easily adapted
for our purpose. To calculate the LDOS, we simply take $|\psi_{0}>$ to
be the vacuum and $A$ be $c_{i\sigma}^{\dag}$. For the spectral function,
we choose $c_{k\sigma}^{\dag}$ instead. The imaginary part of $G_{A}$ then
yields the LDOS and the spectral function respectively.

\section{III. LDOS}

For a homogeneous superconductor, the gap parameter can be directly determined
from the measured LDOS. For an inhomogeneous superconductor, it is not
trivial to invert the LDOS data for $\Delta_{ij}$. Our approach is to try
various $\Delta_{ij}$ configurations to fit the LDOS data. The first set
of data we want to fit are the LDOS spectra measured at representative
points in underdoped cuprates~\cite{el}
 as shown in Figure 1. Motivated by the experimental
gapmap, we consider a cone-shaped distribution~\cite{kap} of the
$\Delta$ order parameter described by the inset of Figure 2. $\Delta$ rises
to $4\Delta_{0}$ at the center of a 400$\times$400 lattice, but it returns
essentially to the background value $\Delta_{0}=0.028 eV$ for distances
larger than ten lattice spacings. The calculated LDOS spectra at a series
of points (at a distance 80,8,6,5,4,2,1 from the origin) are displayed
in Figure 2. They indeed resemble the measured spectra in Figure 1.
In particular, we see that the increase in gapsize
as one moves toward the center of the lattice is accompanied by
a gradual degradation of the coherence peak. As a reference, the LDOS of
a uniform superconductor with $\Delta=4\Delta_{0}$ is included in Figure
2.

One implication of the above result is that the large-gap incoherent
LDOS spectra characteristic of the underdoped cuprates can be interpreted
as a rapid rise of the pairing field in a small region, or a nanoscale
large amplitude fluctuation. There is no need to invoke a certain
unknown charge-ordered zero-temperature
pseudogap state~\cite{el} which competes with DSC.

The other notable feature of Figure 2 (which agrees with Figure 1)
is that the low energy portions of the LDOS spectral are nearly
identical suggesting homogeneous nodal superconductivity coexisting
with the inhomogeneous antinodal feature. Such a contrast between
nodal and antinodal excitations is also reflected in the ARPES
spectra to be discussed in the next section.

Besides the cone-shaped distribution in Figure 2, we have also considered
many other $\Delta$ configurations which do not fit the data so well.
Figure 3 shows the LDOS spectra corresponding to a mesa-like $\Delta$
configuration described in the inset ($\Delta=4\Delta_{0}$ within a
distance of three lattice spacings away from the origin and $\Delta=
\Delta_{0}$ elsewhere).
Such an extended region of high $\Delta$ leads to a higher energy coherent peak in 
the LDOS spectra near the center of the mesa. 
It is quite clear that the data in Figure 1 can discriminate such a 
configuration from the previous configuration of Figure 2.
 While the fit to the data in Figure 1
may not be unique, 
Figure 2 is the best one we have come up with so far.

Although we have considered only a single cone-shaped $\Delta$ configuration,
we anticipate that in real cuprates there would be a disordered array of
cones. As long as the cones do not overlap strongly, the LDOS of the
system should resemble that of a single isolated cone.

The second set of data to fit are the Fourier-Transformed(FT) LDOS.
Due to the constraint of computer time, we again consider only one cone.
The calculated FT-LDOS spectra at various  energies are
displayed in Figure 4 together with the measured one~\cite{el}. As emphasized
by Dell'Anna et al.~\cite{anna},
previous analyses involving impurity scattering
of quasiparticles tend to yield LDOS patterns with extended curve-like
features~\cite{lee,tin,zu,per}
 in the high intensity regions in momentum space. This is in contrast
to the spot-like intensity patterns seen experimentally. 
A zeroth order approximate
calculation by Dell'Anna et al. shows that a mesoscopically
inhomogeneous $\Delta$ distribution indeed yields a  central spot.
Our result in Figure 4 is exact and gives more details than theirs.
In particular, the result exhibits high intensity features along the
diagonal as well as horizontal and vertical directions in good
agreement with experiment. The checkerboard-like intensity modulations
of periodicity about four lattice spacings are also reproduced. 

\section{IV. Spectral Function}

The spatial inhomogeneity revealed by the STM data seems to be at odds
with the well-defined Fermi arc. In addition, the nodal quasiparticle
peak remains well-resolved even in strongly underdoped cuprates. In contrast,
the antinodal quasiparticle peak is well-defined only near optimal
composition~\cite{shen,zhou}.
This contrast has led to the speculation that the nodal and
antinodal excitations have different origins~\cite{el}.
 One is associated with
DSC which dominates in the optimal and overdoped regions, whereas the other
is related to an unknown pseudogap state which competes with DSC and which
dominates in the strongly underdoped samples. Other explanations include
coupling of the electrons with the $(\pi,\pi)$ magnetic excitations~\cite{sc}
and a scattering mechanism operating mainly on the antinodal 
quasiparticles~\cite{zhou}.

We have seen in the previous section that  
 the incoherent LDOS spectra in the underdoped samples can be explained
in terms of inhomogeneous DSC.
Here we attempt to do the same for the ARPES spectra.
 Figure 5 is the calculated spectral density for a disordered
(randomly positioned) array of cones near the nodal direction (5a) and
antinodal direction (5b), both momenta are located on the Fermi surface.
 The nodal quasiparticle peak is indeed well-resolved.
The antinodal peaks are broader, but they are still resolved. This is 
because the $\Delta$ distribution in Figure 5 is dominated by low
$\Delta$ values. 

To simulate the incoherent antinodal spectral functions seen
experimentally in underdoped samples, we fabricate a more disordered
$\Delta$ configuration in Figure 6. The antinodal spectra are indeed
incoherent, whereas the nodal one remains coherent.
We have examined other $\Delta$ configurations, the contrast between
the nodal and antinodal spectral features seems to be generic for
inhomogeneous DSC independent of the details of the inhomogeneity.
Such a result is actually reasonable because the nodal quasiparticles have
vanishingly small excitation energies independent of the magnitude of
$\Delta$, therefore 
they can propagate freely in any inhomogeneous superconductor.
For the opposite reason, the energy of an antinodal quasiparticle is very
sensitive to $\Delta$, high energy antinodal quasiparticles are confined
to regions of high $\Delta$. 
\section{V. Discussion and Conclusion}

For simplicity , we have limited ourselves to only one kind of inhomogeneity
in this paper, the inhomogeneous pairing field. In real cuprates, impurities
are present as well as short-range antiferromagnetism. They could also
affect the LDOS~\cite{atk,yeh}
and  the spectral weight function. Further theoretical study is required
to include those effects.

Despite the limitation,
 our results so far support the following conclusions: (1)
Inhomogeneous DSC is an important determining factor in the LDOS and
ARPES spectra; (2) It can explain many unusual features of the experimental
spectra without invoking an unknown state which competes with DSC,
at least for optimally doped and moderately underdoped systems.
The "dichotomy" between nodal and antinodal excitations seems to be
a mere consequence of the inhomogeneous  d-wave pairing field.

This work was partially supported by the Texas Center for Superconductivity,
the Robert A. Welch Foundation (grant number E-1070),
and the National Science Council of Taiwan under contract number NSC
92-2112-M-110-006. We thank Degang Zhang and Hongyi Chen for useful
conversation.One of the authors (WPS) thanks S. F. Tsay and 
 the Deparment of Physics
at the National Sun Yat-Sen University for their hospitality during summer
2004.

Fig 1: LDOS spectra measured at representative points in underdoped cuprates,
data taken from ref.~\cite{el}.

Fig 2: Calculated LDOS spectra for the spiky $\Delta$ configuration described
in the inset. The curve labeled `uniform $\Delta$' is included as a
reference. It is the LDOS spectrum for a uniform $\Delta=4\Delta_{0}$.

Fig 3: Calculated LDOS spectra(at a distance 43,4,3,2,1,0 from the origin) for
the $\Delta$ configuration described in the inset.

Fig 4: Calculated FT-LDOS (left column) for a cone-shaped $\Delta$ 
distribution at various biases ($\omega=-8,-16,-26$ meV from top to bottom) compared
with data (right column) taken from ref. ~\cite{el}.

Fig 5: Spectral functions along the nodal and antinodal
directions corresponding to the $\Delta$ configuration on the left panel.

Fig 6: Same as Figure 5 for a different $\Delta$ distribution.

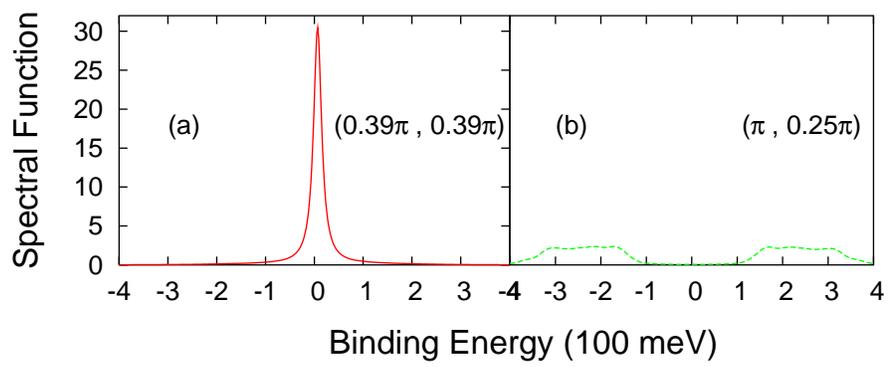
\end{document}